\documentclass[letterpaper,floatfix]{quantumarticle}
\pdfoutput=1
\usepackage[utf8]{inputenc}
\usepackage[english]{babel}
\usepackage[T1]{fontenc}
\usepackage{amsmath,graphicx}
\usepackage{hyperref}
\usepackage[numbers]{natbib}
\usepackage{color}
\usepackage[dvipsnames]{xcolor}
\usepackage{mathtools}
\usepackage{multirow}
\usepackage{tikz}
\usepackage{hyperref}
\hypersetup{colorlinks=true, citecolor=blue, linkcolor=blue}
\usepackage{url}

\makeatletter
\DeclareFontFamily{OMX}{MnSymbolE}{}
\DeclareSymbolFont{MnLargeSymbols}{OMX}{MnSymbolE}{m}{n}
\SetSymbolFont{MnLargeSymbols}{bold}{OMX}{MnSymbolE}{b}{n}
\DeclareFontShape{OMX}{MnSymbolE}{m}{n}{
    <-6>  MnSymbolE5
   <6-7>  MnSymbolE6
   <7-8>  MnSymbolE7
   <8-9>  MnSymbolE8
   <9-10> MnSymbolE9
  <10-12> MnSymbolE10
  <12->   MnSymbolE12
}{}
\DeclareFontShape{OMX}{MnSymbolE}{b}{n}{
    <-6>  MnSymbolE-Bold5
   <6-7>  MnSymbolE-Bold6
   <7-8>  MnSymbolE-Bold7
   <8-9>  MnSymbolE-Bold8
   <9-10> MnSymbolE-Bold9
  <10-12> MnSymbolE-Bold10
  <12->   MnSymbolE-Bold12
}{}

\let\llangle\@undefined
\let\rrangle\@undefined
\DeclareMathDelimiter{\llangle}{\mathopen}%
                     {MnLargeSymbols}{'164}{MnLargeSymbols}{'164}
\DeclareMathDelimiter{\rrangle}{\mathclose}%
                     {MnLargeSymbols}{'171}{MnLargeSymbols}{'171}
\makeatother

\begin{document}

\newcommand{\cE}{\mathcal{E}}
\newcommand{\cL}{\mathcal{L}}
\newcommand{\cB}{\mathcal{B}}
\newcommand{\cH}{\mathcal{H}}

\newcommand{\expect}[1]{\ensuremath{\left\langle#1\right\rangle}}
\newcommand{\ket}[1]{\ensuremath{\left|#1\right\rangle}}
\newcommand{\bra}[1]{\ensuremath{\left\langle#1\right|}}
\newcommand{\braket}[2]{\ensuremath{\left\langle#1|#2\right\rangle}}
\newcommand{\ketbra}[2]{\ket{#1}\!\!\bra{#2}}
\newcommand{\braopket}[3]{\ensuremath{\bra{#1}#2\ket{#3}}}
\newcommand{\proj}[1]{\ketbra{#1}{#1}}
\newcommand{\sket}[1]{\ensuremath{\left|#1\right\rrangle}}
\newcommand{\sbra}[1]{\ensuremath{\left\llangle#1\right|}}
\newcommand{\sbraket}[2]{\ensuremath{\left\llangle#1|#2\right\rrangle}}
\newcommand{\sketbra}[2]{\sket{#1}\!\!\sbra{#2}}
\newcommand{\sbraopket}[3]{\ensuremath{\sbra{#1}#2\sket{#3}}}
\newcommand{\sproj}[1]{\sketbra{#1}{#1}}
\def\Id{1\!\mathrm{l}}
\newcommand{\Tr}{\mathrm{Tr}}

\title{Easy better quantum process tomography}

\author{Robin Blume-Kohout}
\affiliation{Quantum Performance Laboratory\\Sandia National Laboratories, Albuquerque, New Mexico 87185}
\email{robin@blumekohout.com}
\author{Kenneth Rudinger}
\affiliation{Quantum Performance Laboratory\\Sandia National Laboratories, Albuquerque, New Mexico 87185}

\author{Timothy Proctor}
\affiliation{Quantum Performance Laboratory\\Sandia National Laboratories, Livermore, California 94550}

\begin{abstract}
Quantum process tomography (QPT), used to estimate the linear map that best describes a quantum operation, is usually performed using \emph{a priori} assumptions about state preparation and measurement (SPAM), which yield a biased and inconsistent estimator.  This estimate can be made more accurate and less biased by incorporating SPAM-calibration data.  Unobservable properties of the SPAM operations introduce a small gauge freedom that can be regularized using the \emph{a priori} SPAM.  We give an explicit correction procedure for standard linear-inversion QPT and overcomplete linear-inversion QPT, and describe how to extend it to statistically principled estimators like maximum likelihood estimation.
\end{abstract}

\maketitle

Quantum process tomography (QPT) \cite{Chuang1997-dj, O-Brien2004-ht,Hashim2024-om} is used to estimate the linear map that best describes a quantum operation.  Standard QPT is vulnerable to state preparation and measurement or ``SPAM'' \cite{Magesan2012-dz} error \cite{Quesada2013-th,Merkel2013-lx,Stark2014-gk,Blume-Kohout2013-ry,Greenbaum2015-mp}.  Unless the tomographer knows and uses \emph{exactly} correct models for the initial states and final measurements used to probe the unknown operation, QPT will yield biased, inaccurate estimates.  Gate set tomography (GST) \cite{Merkel2013-lx,Greenbaum2015-mp,Blume-Kohout2017-no,Nielsen2021-nu} was developed to eliminate this problem, but because GST requires more experimental effort and more complicated analysis than QPT, it is sometimes viewed as ``too much work''.  QPT is still commonly used despite its known flaws.

This paper shows how to eliminate QPT's SPAM-induced inaccuracy with minimal extra effort.  This ``better QPT'' algorithm requires only $2\times$ as much experimental data and is almost as simple as standard QPT.  It uses the essential idea behind GST but applies it to estimate a \emph{single} operation.  It acknowledges the gauge freedom \cite{Blume-Kohout2017-no,Nielsen2021-nu,Proctor2017-wc,Rudnicki2018-qo,Lin2019-qx, Chen2023-lo} associated with GST (and ignored by QPT), but deals with it in the simplest possible way.  If the \emph{a priori} SPAM operations were rank-1 (perfect), then the corrected estimate will be not just more accurate, but (truthfully) higher-fidelity.

\section{Standard QPT}

QPT estimates an unknown operation $G$ on a quantum system with $d$-dimensional Hilbert space $\cH$ by
\begin{enumerate}
\item choosing a set of known quantum states, described by $d\times d$ density matrices $\{\overline{\rho}_j\}$, to which the unknown operation will be applied,
\item choosing a set of known quantum measurements, described by positive operator-valued measures (POVMs) with outcomes described by $d\times d$ positive semi-definite ``effects'' $\{\overline{E}_i\}$, to perform on the transformed states,
\item repeatedly preparing each $\overline{\rho}_j$, applying $G$, and performing each of the various measurements, so that, for every $(i,j)$, the probability $P_{i,j}$ of observing effect $\overline{E}_i$ given $G[\overline{\rho}_j]$ can be estimated to reasonable precision.
\end{enumerate}
The resulting data can be analyzed by representing each density matrix $\overline{\rho}_j$ as a column vector $\sket{\overline{\rho}_j}$ in the $d^2$-dimensional Hilbert-Schmidt space $\cB(\cH)$ of operators and each effect $\overline{E}_i$ as a row vector $\sbra{\overline{E}_i}$ in $\cB(\cH)$, and then arranging the probabilities $P_{i,j}$ into a matrix that can be written using Born's rule ($\mathrm{Pr}(E|\rho) = \Tr[E\rho]$) and the Hilbert-Schmidt inner product ($\sbraket{A}{B} = \Tr[A^\dagger B]$) as
\begin{equation}
P_{i,j} \equiv \mathrm{Pr}\left( \overline{E}_i | G[\overline{\rho}_j] \right) = \sbraopket{\overline{E}_i}{G}{\overline{\rho}_j}. \nonumber
\end{equation}
An easy and elegant way to write this is by constructing two matrices,
\begin{eqnarray}
M_0 &=& \left(\begin{array}{c} \sbra{\overline{E}_1} \\ \sbra{\overline{E}_2} \\ \vdots \end{array}\right), \nonumber\\
S_0 &=& \left(\sket{\overline{\rho}_1} \sket{\overline{\rho}_2} \hdots \right), \nonumber
\end{eqnarray}
so that
\begin{equation} \label{eq:PMGS}
P = M_0 G S_0.
\end{equation}
Now, if $\hat{P}$ contains the estimated probabilities, then the ``linear inversion'' estimate of the unknown process is simply
\begin{equation}
\hat{G}_0 = M_0^{-1} \hat{P} S_0^{-1}. \label{eq:QPT}
\end{equation}
More sophisticated estimators like maximum likelihood estimation \cite{Hradil1997-tg, Banaszek1999-eh, James2001-aa, Wasserman2013-fx} tweak this simple algorithm to mitigate the effects of significant finite-sample fluctuations in $\hat{P}$, which include violation of the complete positivity constraint $\left(\hat{G}\otimes\Id\right)[\proj{\mathrm{\textsc{bell}}}]\geq 0$ and inconsistency between probabilities associated with ``overcomplete'' sets of states and/or effects.

\section{The problem with QPT}

QPT works fine if the $\{\overline{\rho}_j\}$ and $\{\overline{E}_i\}$ do, in fact, represent the true states and effects used in the QPT experiment.  But most QPT experiments are analyzed under the implausible assumption that the state preparation and measurement (SPAM) operations -- the \emph{true} $\{\rho_j\}$ and $\{E_i\}$ -- are exactly equal to their respective rank-1 states $\{\overline{\rho}_j\}$ and effects $\{\overline{E}_i\}$ describing the experimenter's intent.  This model is obviously wrong, because quantum operations are always subject to errors. Assuming perfect SPAM yields incorrect $M_0$ and $S_0$ matrices that (in turn) yield incorrect and potentially misleading estimates $\hat{G}_0$.  

Some QPT experiments do better by attempting to model the SPAM operations that make up $M_0$ and $S_0$ as realistically noisy operations, inferred from data.  However, the canonical way to estimate a density matrix is quantum state tomography, which relies on known POVM effects\ldots and the canonical way to estimate a POVM is quantum measurement tomography, which relies on known states.  This creates an obvious circularity, to which no really satisfactory solution is known.  It is now understood that states and measurements literally \emph{can't} be fully learned or estimated, because controllable quantum systems have a gauge symmetry \cite{Blume-Kohout2017-no, Nielsen2021-nu,Proctor2017-wc,Rudnicki2018-qo,Lin2019-qx, Chen2023-lo}.

In principle, this gauge symmetry creates existential problems for state, measurement, and process tomography.  Yet all three of these protocols are still regularly performed and their results published, suggesting the gauge problem is not as unsolvable in practice as it is in principle.

\section{Better QPT}

QPT can be improved by learning about the SPAM operations and adjusting the estimate $\hat{G}$ with that knowledge.  This can be done by performing an additional ``process tomography on nothing'' experiment to estimate each of the probabilities
\begin{equation}
I_{i,j} \equiv \mathrm{Pr}( E_i | \rho_j ) = \sbraket{E_i}{\rho_j}.\nonumber
\end{equation}
The unknown SPAM operations are now modeled by defining two \emph{unknown} matrices
\begin{eqnarray}
M &=& \left(\begin{array}{c} \sbra{E_1} \\ \sbra{E_2} \\ \vdots \end{array}\right), \nonumber\\
S &=& \left(\sket{\rho_1} \sket{\rho_2} \hdots \right),\nonumber
\end{eqnarray}
of which $M_0$ and $S_0$ are just \emph{a priori} estimates.  There are now two matrices of observable probabilities,
\begin{eqnarray}
I &=& MS, \nonumber\\
P &=& MGS. \nonumber
\end{eqnarray}
If the \emph{a priori} guesses about the $\{\rho_j\}$ and $\{E_i\}$ were correct -- i.e., $M = M_0$ and $S = S_0$, then taking data and estimating the elements of $I$ will yield $\hat{I} = M_0S_0$ to within finite-sample fluctuations (error bars).  If this is observed, the assumptions of standard QPT are validated and Eq.~\ref{eq:QPT} can be used directly.

Otherwise, $\hat{I}$ can be used to construct a $\hat{G}$ that is more accurate than $\hat{G}_0$.  Writing
\begin{equation}
I = M_0 \cE S_0,\nonumber
\end{equation}
defines a \emph{SPAM error superoperator} $\cE$ that captures the SPAM operations' deviation from expectations.  It can be estimated as 
\begin{equation}
\hat{\cE} = M_0^{-1} \hat{I} S_0^{-1}.\nonumber
\end{equation}
Note that $\cE$ is not required to be a completely positive trace-preserving map; it merely provides an effective description of the errors in the SPAM operations.  

If the states $\{\rho_j\}$ were known to be perfect ($S=S_0$), then the rows of $M = M_0 \cE$ would reveal the noisy measurement effects $\{E_i\}$.  Similarly, if the effects $\{E_i\}$ were known to be perfect ($M=M_0$) then the columns of $S = \cE S_0$ would reveal the noisy states $\{\rho_j\}$.  

In practice, both are noisy. The data cannot tell us how to divide the SPAM error described by $\hat{\cE}$ between $M$ (measurements) and $S$ (states).  In principle, any factorization $\hat{\cE}=\alpha \beta$ into invertible matrices $\alpha$ and $\beta$ yields a valid theory about the SPAM operations:
\begin{eqnarray}
\hat{S} &=& \beta S_0, \nonumber\\
\hat{M} &=& M_0 \alpha = M_0 \cE \beta^{-1},\nonumber
\end{eqnarray}
and in turn a valid theory about $G$:
\begin{eqnarray}
\hat{G} &=& \hat{M}^{-1}\hat{P} \hat{S}^{-1} \nonumber\\
	&=& \alpha^{-1} M_0^{-1} \hat{P} S_0^{-1} \beta^{-1} \nonumber\\
	&=& \alpha^{-1} \hat{G}_0 \beta^{-1} \nonumber\\
	&=& \beta \hat{\cE}^{-1} \hat{G}_0 \beta^{-1}.\nonumber
\end{eqnarray}
But this constitutes excessive respect for gauge freedom.  We should respect the \emph{a priori} estimates $M \approx M_0$ and $S \approx S_0$, while adjusting them to respect the data.  By choosing $\beta = \hat{\cE}^{1/2}$, we can divide the SPAM error $\hat{\cE}$ equally between states and effects:
\begin{eqnarray}
\hat{S} &=& \hat{\cE}^{1/2} S_0 \nonumber\\
\hat{M} &=& M_0 \hat{\cE}^{1/2}  \nonumber\\
&\Rightarrow& \hat{G} = \hat{\cE}^{-1/2} \hat{G}_0 \hat{\cE}^{-1/2}. \label{eq:betterQPT}
\end{eqnarray}
If there is reason to believe that either state preparation or measurement is more error-prone, then it is equally valid to divide the SPAM error unequally by choosing $\beta = \hat{\cE}^{p}$ for some $p \in [0,1]$ so that:
\begin{eqnarray}
\hat{S} &=& \hat{\cE}^{p} S_0 \nonumber\\
\hat{M} &=& M_0 \hat{\cE}^{1-p}  \nonumber\\
&\Rightarrow& \hat{G}(p) = \hat{\cE}^{p-1} \hat{G}_0 \hat{\cE}^{-p}.\nonumber
\end{eqnarray}
If the observed SPAM error $\hat{\cE}$ commutes with $\hat{G}_0$ -- e.g., if it is a depolarizing channel -- then these options all coincide.

\begin{figure}[t!]
    \centering  
   \includegraphics[width=8cm]{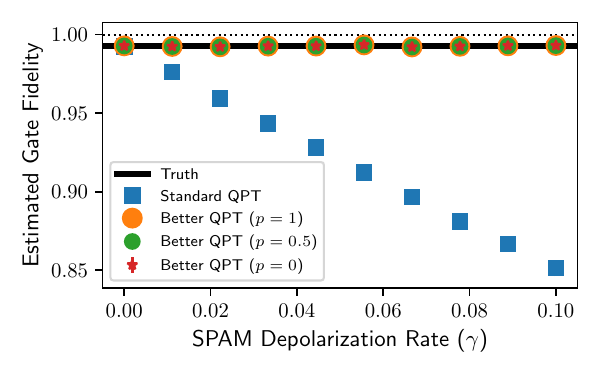}
   
   \vspace{-0.05in}
\includegraphics[width=8cm]{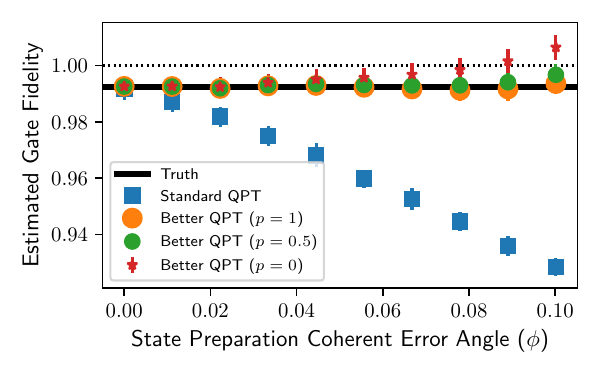}

   \vspace{-0.05in}
\includegraphics[width=8cm]{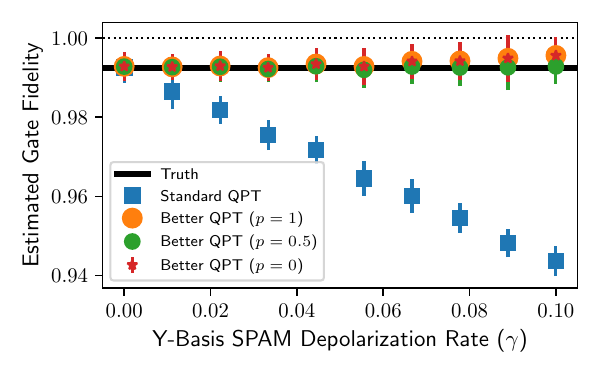}
    \caption{\textbf{Comparative accuracy of standard and SPAM-corrected QPT:} We compared the accuracy of standard QPT (blue) and SPAM-adjusted QPT (orange, green, red) by simulating both protocols on an $X_{\pi/2}$ gate with $1\%$ depolarizing noise, with 5000 shots per circuit (see main text for details).  To quantify accuracy, we used the process matrix estimated by each protocol to estimate the noisy gate's fidelity with a noiseless $X_{\pi/2}$ unitary, and plotted this estimated fidelity as a function of the strength of depolarizing SPAM noise (top panel), coherent SPAM noise (middle panel; see main text for details), and depolarizing errors on only $Y$-basis SPAM (bottom panel).  When performing SPAM-corrected QPT, we assigned a fraction $p=0$ (orange), $p=0.5$ (green), or $p=1$ (red) of the observed SPAM error to state preparation, and the remainder to measurement.  This gauge choice had a weak impact on accuracy for coherent SPAM and Y-basis depolarization errors, but no effect for depolarizing SPAM error.}
    \label{fig:simulations}
\end{figure}

\section{Simulations}

To demonstrate the benefits of this protocol, we simulated standard and better QPT with SPAM errors on a single-qubit $X_{\pi/2}$ gate with $1\%$ uniform depolarizing noise (i.e., an ideal $X_{\pi/2}$ unitary followed by the error process $\mathcal{D}_{\gamma}[\rho] = (1-\gamma) \rho + \gamma\Id/2$ with $\gamma = 0.01$).  In the ideal experiment, both the input states and the measured effects were given by Pauli eigenstates $\{\proj{+_x},\proj{-_x},\proj{+_y},\proj{+_z}\}$.  We added three kinds of SPAM noise:
\begin{enumerate}
\item \textbf{Depolarizing}: we applied uniform depolarization ($\mathcal{D}_{\gamma}$), with varying strength $\gamma$, to all state preparations and measurements. The total SPAM depolarization factor is $(1-\gamma)^2 \approx 1-2\gamma$, since both states and measurements are depolarized. 
\item \textbf{Coherent}: measurements are ideal and error-free, but each ideal input state $\ket{\psi}$ is replaced with $\ket{\psi_\phi} = \cos\phi\ket{\psi} + \sin\phi\ket{\psi_{\perp}}$ where $\ket{\psi_{\perp}}$ is the unique orthogonal state and $\phi$ is varied.
\item \textbf{Y-basis depolarizing}: we applied depolarization ($\mathcal{D}_{\gamma}$) with varying strength $\gamma$ to \textit{only} the $Y$-basis state preparation and measurement.  $X$- and $Z$-basis SPAM are error-free. This model produces a non-CP $\cE$, and might be encountered in a surface-code logical qubit where $Y$ prep/measure is hard.
\end{enumerate}

We simulated 50 runs of QPT experiments with 5000 shots of each circuit (12 circuits for standard QPT, 24 for SPAM-corrected QPT).  All error bars are $1\sigma$ and derived from the variance of the 50 repetitions.  Because QPT is often used to estimate the fidelity of an experimental gate with its ideal target, we quantify tomographic accuracy by comparing the value of that fidelity computed from the QPT estimate to its \emph{true} value.  This comparison is shown in Figure~\ref{fig:simulations}, for three SPAM error models, as a function of SPAM error magnitude ($\gamma$ or $\phi$). Accuracy depends on the tomographer's \emph{a priori} division of SPAM error between state preparation and measurement ($p$), so we illustrate three choices that assign $p=0$, $p=50\%$, and $p=100\%$ of the SPAM error (respectively) to state preparation. 

SPAM-corrected QPT is more accurate than standard QPT in every case.  For depolarizing SPAM error (top panel), the parameter $p$ that determines how SPAM error is divided between between state preparation and measurement has no effect, but for coherent errors (middle) and state-dependent depolarization (bottom), guessing wrong about the division of error causes a small degradation in accuracy.

\begin{figure}[t!]
    \centering  
   \includegraphics[width=8cm]{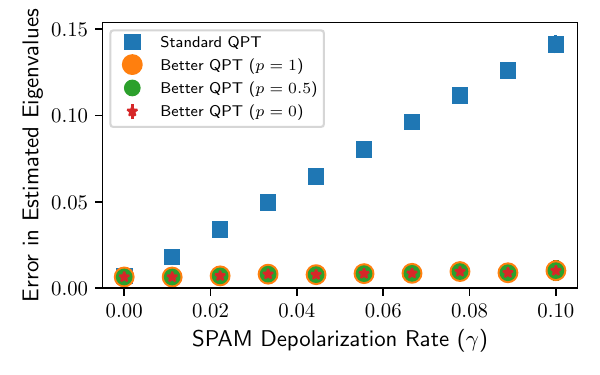}
\includegraphics[width=8cm]{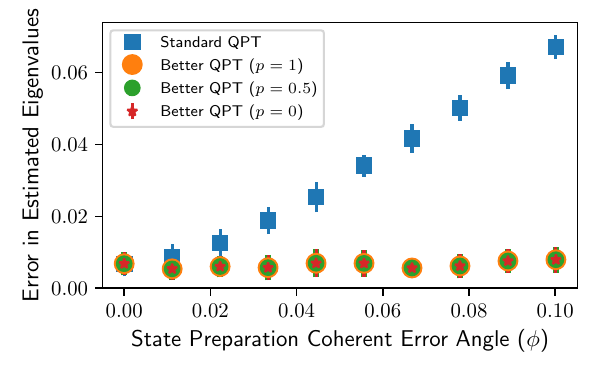}
\includegraphics[width=8cm]{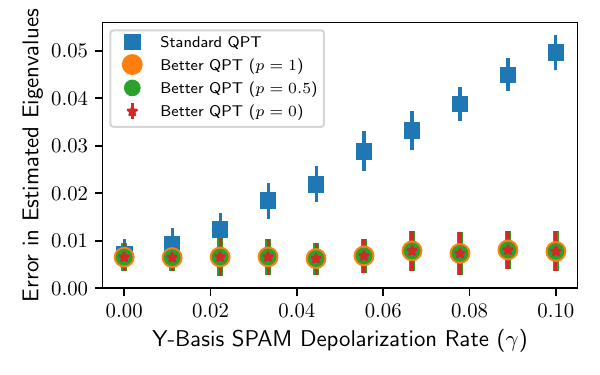}
    \caption{\textbf{Error in estimated gate eigenvalues:} Gauge-invariant quantities like the eigenvalues of the estimated superoperator (transfer matrix) should be independent of how SPAM error is divided.  We analyzed the results of simulated standard and SPAM-corrected QPT to determine how inaccurately they estimated gate eigenvalues, using the figure of merit $\delta = \frac{1}{4}\sum_i|\lambda_i - \tilde{\lambda}_i|$, where $\lambda_i$ are the true eigenvalues and $\tilde{\lambda}_i$ the estimates.  Standard QPT's inaccuracy grows with SPAM error, while SPAM-corrected QPT's accuracy is unaffected by SPAM error.}
    \label{fig:simulations2}
\end{figure}

The division of SPAM error is a choice of gauge.  Figure~\ref{fig:simulations} shows that even when a gauge-variant quantity (fidelity) is computed, this gauge choice has a relatively minor impact when the SPAM error is small\footnote{Careful observation reveals that the estimated gate fidelity can be greater than 1 for large coherent SPAM errors.  This is purely an effect of guessing the wrong gauge, and could in principle be avoided by numerically constraining the gauge choice to enforce complete positivity of $\hat{G}$. But this is outside the scope of SPAM-corrected QPT, whose philosophy is to trust the \textit{a priori} SPAM as much as possible.}.  But if we focus on gauge-invariant properties -- e.g., the eigenvalues of the estimated process matrix -- the impact of this gauge choice should vanish entirely.  Figure~\ref{fig:simulations2} confirms that SPAM-corrected QPT estimates gate eigenvalues very accurately, and (unlike standard QPT) is completely immune to SPAM error.  We quantified error in the estimated eigenvalues by $\delta = \frac{1}{4}\sum_i|\lambda_i - \tilde{\lambda}_i|$ where $\lambda_i$ are the true eigenvalues and $\tilde{\lambda}_i$ the estimates.  We also found that the residual error in the eigenvalues declines to zero as the number of shots is increased (not shown).

\section{Overcomplete data}

The derivation above assumed the use of states and measurements that are informationally complete, not overcomplete.  If $K > d^2$ states and/or effects are considered, then the matrices $M$ and/or $S$ become rectangular, and do not have inverses.  In standard QPT, Eq.~\ref{eq:PMGS} ($P = M_0 G S_0$) becomes overconstrained and has no solutions, because the rank of $M_0 G S_0$ is at most $d^2$, while $P$ can have rank $K$.

The easiest way to deal with this is with ordinary least-squares (OLS) fitting, finding the $\hat{G}$ that minimizes $\|M_0 G S_0 - P\|_2^2$.  This admits a closed-form solution, given by replacing matrix inverses with Moore-Penrose pseudoinverses, e.g.,
\begin{eqnarray}
M_0^+ &\equiv& (M_0^\dagger M_0)^{-1} M_0^\dagger \nonumber\\
S_0^+ &\equiv& S_0^\dagger (S_0 S_0^\dagger)^{-1}.\nonumber
\end{eqnarray}
The standard OLS QPT estimate of $G$ for overcomplete data is therefore
\begin{equation}
\hat{G}_0 = M_0^+ \hat{P} S_0^+.\nonumber
\end{equation}
This estimate can also be SPAM-corrected using the information obtained by measuring $I = MS$.  However, the procedure (and its derivation) are slightly more complicated and illuminating.

Since $M$ and $S$ are (respectively) $K\times d^2$ and $d^2 \times K$ matrices, the $K \times K$ matrix of empirical probabilities $I = MS$ has rank at most $d^2 < K$.  However, an experimentally estimated $\hat{I}$ will generally be full-rank because of (1) sampling error in the empirical probabilities, and/or (2) inadvertent manipulation of a larger state space than intended (a form of non-Markovian error).  If $\mathrm{rank}(\hat{I}) > d^2$, then the only way to find $M$ and $S$ satisfying $MS=\hat{I}$ is by ``truncating'' $\hat{I}$ to the closest rank-$d^2$ matrix.  This can be done by performing a singular value decomposition and replacing $\hat{I}$'s smallest $K-d^2$ singular values with 0.

Truncation also improves accuracy, under the assumption that deviations from rank-$d^2$ are caused by finite-sample fluctuations\footnote{If such deviations result from manipulating a larger state space than intended, then this ansatz does not hold -- but in this case the entire tomographic problem is ill-posed because there is no ``true'' $d^2\times d^2$ process to be estimated as accurately as possible.}.  For this reason, it is advantageous (though not required) to truncate $\hat{P}$ as well.

Once truncated, $\hat{I}$ can be factored as $\hat{I} = MS$ for properly sized $M$ and $S$.  But $M$ and $S$ are not unique in this factorization.  If $MS=\hat{I}$ then $(MQ)(Q^{-1}S) = \hat{I}$ as well, for any invertible $d^2\times d^2$ matrix $Q$.  We want one for which $M \approx M_0$ and $S \approx S_0$.  We could minimize $\|M - M_0\|_2$ by choosing
\begin{eqnarray}
S_{\mathrm{m-opt}} &\equiv& M_0^+ \hat{I}, \label{eq:28} \\
M_{\mathrm{m-opt}} &\equiv& \hat{I} S_{\mathrm{m-opt}}^+. 
\end{eqnarray}
This is a valid factorization -- i.e., $M_{\mathrm{m-opt}}S_{\mathrm{m-opt}} = \hat{I} S_{\mathrm{m-opt}}^+S_{\mathrm{m-opt}} = \hat{I}$ --  because (1) $S_{\mathrm{m-opt}}^+S_{\mathrm{m-opt}}$ is the product of a rank-$d^2$ matrix and its pseudo-inverse and must therefore be a rank-$d^2$ projector, and (2) $S_{\mathrm{m-opt}}^+S_{\mathrm{m-opt}} = S_{\mathrm{m-opt}}^+ M_0^+ \hat{I}$ necessarily annihilates any vector in the orthogonal complement of $\hat{I}$'s ($d^2$-dimensional) row space.  Therefore, $S_{\mathrm{m-opt}}^+S_{\mathrm{m-opt}}$ must be the projector onto the row space of $\hat{I}$.\footnote{Note that pseudoinverses do not obey the same rules as true inverses, so in general $(AB)^+ \neq B^+A^+$.  However, such identities often hold \emph{approximately}.  If $S_{\mathrm{m-opt}} \approx \hat{I}^+ M_0$, then $M_{\mathrm{m-opt}} \approx \hat{I} \hat{I}^+ M_0 = \Pi_c M_0$, where $\Pi_c \equiv \hat{I}\hat{I}^{+}$ is the projector onto $\hat{I}$'s column space.  Approximations of this form can yield useful intuitions.}

Alternatively, we could minimize $\|S-S_0\|_2$ by choosing
\begin{eqnarray}
M_{\mathrm{s-opt}} &\equiv& \hat{I} S_0^+, \\
S_{\mathrm{s-opt}} &\equiv& M_{\mathrm{s-opt}}^+ \hat{I}.
\end{eqnarray}
To split the difference as in the previous section, we can write
\begin{equation}
\hat{I} = M_{\mathrm{m-opt}} \hat{\cE} S_\mathrm{s-opt}, \nonumber
\end{equation}
where $\hat{\cE}$ is once again a $d^2 \times d^2$ matrix\footnote{In the previous section, $\hat{\cE}$ described \emph{all} the SPAM error.  Here, it describes only the SPAM error that can be ``sloshed'', by choice of gauge, between $M$ and $S$.  In contrast, the discrepancy between $M_0$ and $M_{\mathrm{m-opt}}$ (or $M_0$ and $M_{\mathrm{m-opt}}$) is objective and intrinsic to $M$ (or $S$), because \emph{no} feasible estimate eliminates it.}.  We can then extract $\hat{\cE}$ as
\begin{equation}
\hat{\cE} = M_{\mathrm{m-opt}}^+ \hat{I} S_\mathrm{s-opt}^+, \label{eq:33}
\end{equation}
and obtain good estimates of $M$ and $S$ as
\begin{eqnarray}
\hat{M} = M_{\mathrm{m-opt}}\hat{\cE}^{1/2}, \label{eq:Mhat}\\
\hat{S} = \hat{\cE}^{1/2}S_{\mathrm{s-opt}}. \label{eq:Shat}
\end{eqnarray}
Finally, using these estimates of $M$ and $S$, we can analyze the process tomography data ($P$) as
\begin{equation}
\hat{G} = \hat{M}^+ \hat{P} \hat{S}^+. \label{eq:overcompleteQPT}
\end{equation}
It is possible to expand Eq.~\ref{eq:overcompleteQPT} using the definition of the pseudoinverse and Eqs.~\ref{eq:Mhat}-\ref{eq:Shat} and their predecessors (Eqs.~\ref{eq:28}-\ref{eq:33}), but the nested pseudoinverses don't simplify.  However, applying the uncontrolled approximation $(AB)^+ \approx B^+A^+$, yields the suggestive approximation
\begin{equation}
\hat{G} \approx \hat{\cE}^{-1/2} M_0^+ \Pi_c \hat{P} \Pi_r S_0 \hat{\cE}^{-1/2},\nonumber
\end{equation}
where $\Pi_c$ and $\Pi_r$ project onto $\hat{I}$'s column and row spaces, respectively.  This mirrors Eq.~\ref{eq:betterQPT}, but with the addition of projections onto the most active subspaces of $\hat{I}$ (and optionally also of $\hat{P}$).

\section{Principled statistical estimation}

The analysis above equates ``estimation'' with linear regression via ordinary least squares.  This is an oversimplification, although a very useful one.  The estimators in Eqs.~\ref{eq:betterQPT} and \ref{eq:overcompleteQPT} can absolutely be used in practice, but they will be suboptimal for at least three reasons.
\begin{enumerate}
\item In Eq.~\ref{eq:overcompleteQPT}, the matrix of empirical probabilities for QPT ($\hat{P}$) is projected onto the support of $\hat{I}$ by $\hat{I}^+ \hat{I}$, because in a previous step we used $\hat{I}$ to identify the active subspaces supporting $M$ and $S$. Since $K>d^2$ states or effects in a $d^2$-dimensional space cannot be linearly independent, \emph{some} subspace restriction is necessary.  But $\hat{P}$ also contains information about active subspaces, which is not used because $\hat{I}$ and $\hat{P}$ are analyzed separately.
\item The estimated probabilities in $\hat{I}$ and $\hat{P}$ have varying precisions, because the Fisher information \cite{ly2017tutorial} of multinomial parameters varies with the parameter.  Ordinary least squares ignores this.
\item The process estimate $\hat{G}$ and/or the SPAM estimates $\hat{M}$ and $\hat{S}$ may violate physical constraints (each $\rho_i$ and $E_j$ must be positive semidefinite, and $G$ must be completely positive and trace-preserving, or CPTP).  These constraints are nontrivial, and cannot usually be enforced in closed form (except in the special case of Ref.~\cite{Smolin2012-mm}).
\end{enumerate}
These deficiencies can all be addressed by replacing closed-form linear regression with numerical algorithms for statistical estimation that fit unknown parameters to data using statistical principles such as maximum likelihood estimation (MLE)  \cite{Hradil1997-tg,Banaszek1999-eh,James2001-aa,Wasserman2013-fx}.  Principled statistical estimators typically yield more accurate results than least-squares regression.

There are at least three ways to integrate sophisticated statistical estimators with the ``better QPT'' workflow described above.
\begin{enumerate}
\item \textbf{Independent estimation of $\cE$ and $G_0$}:  Use a statistical principle (e.g. MLE) to estimate SPAM error $\hat{\cE}$ from data in $\hat{I}$ and to estimate $\hat{G}_0$ from data in $\hat{P}$, and then compute $\hat{G}$ using Eq.~\ref{eq:betterQPT} or \ref{eq:overcompleteQPT}.  This is the easiest approach, but relative n\"aive -- e.g., $\hat{G}$ may not satisfy positivity.
\item \textbf{Sequential estimation of SPAM and $G$}:  Use a statistical principle (e.g. MLE) to estimate $\hat{M}$ and $\hat{S}$ from data in $\hat{I}$, and then to estimate $\hat{G}$ from the data in $\hat{P}$ given the previously estimated $\hat{M}$ and $\hat{S}$.  This approach can guarantee positivity of $\hat{G}$, but does not make optimal use of data -- e.g., the active subspaces for $\hat{M}$ and $\hat{S}$ are not informed by data in $\hat{P}$.
\item \textbf{Joint estimation of SPAM and $G$}:  Use a statistical principle (e.g. MLE) to simultaneously estimate $\hat{M}$, $\hat{S}$, and $\hat{G}$ from the data in $\hat{I}$ and $\hat{P}$.  This is conceptually elegant and makes optimal use of data \emph{and} constraints, but requires numerical maximization or integration of a likelihood function $\cL(\hat{M},\hat{S},\hat{G}) = \mathrm{Pr}( \hat{I},\hat{P} | \hat{M},\hat{S},\hat{G} )$ that will not generally be convex.  Note that in this approach, $\hat{M}$ and $\hat{S}$ are essentially ``nuisance parameters'' -- they are not needed for the final product ($\hat{G}$), but need to be estimated jointly to remove bias from $\hat{G}$.
\end{enumerate}
The comparative performance of these three approaches -- e.g., whether the first is good enough, or whether the third provides meaningful advantage over the second -- is an interesting question beyond the scope of this paper.

\section*{Acknowledgements}
This material was funded in part by the U.S. Department of Energy, Office of Science, Office of Advanced Scientific Computing Research, Quantum Testbed Pathfinder Program. This article has been coauthored by employees of National Technology \& Engineering Solutions of Sandia, LLC under Contract No.~DE-NA0003525 with the U.S.~Department of Energy (DOE). The employees own all right, title, and interest in and to the article and are solely responsible for its contents. The U.S.~Government retains, and the publisher, by accepting the article for publication, acknowledges that the U.S.~Government retains, a non-exclusive, paid-up, irrevocable, worldwide license to publish or reproduce the published form of this article or allow others to do so, for U.S.~Government purposes. The DOE will provide public access to these results of federally sponsored research in accordance with the DOE Public Access Plan \href{https://www.energy.gov/downloads/doe-public-access-plan}{https://www.energy.gov/downloads/doe-public-access-plan}.  This paper describes objective technical results and analysis. Any subjective views or opinions that might be expressed in the paper do not necessarily represent the views of the U.S.~Department of Energy or the U.S.~Government.

\bibliographystyle{quantum}
\bibliography{Bibliography}
\end{document}